\documentclass[11pt]{article}%
\usepackage{amsfonts}
\usepackage{amsmath}
\usepackage{amssymb}
\usepackage{graphicx}%
\providecommand{\U}[1]{\protect\rule{.1in}{.1in}}

\begin{document}

\title{A Geometric Picture of the Wave Function:\ Fermi's Trick}
\author{Maurice A. de Gosson\thanks{maurice.de.gosson@univie.ac.at}\\University of Vienna\\Faculty of Mathematics, NuHAG\\Nordbergstr. 15, 1090 Vienna\\AUSTRIA}
\maketitle

\begin{abstract}
We show that there is a one-to-one correspondence between wave functions and
surfaces in the position-momentum phase plane bounded by a closed curve
satisfying an exact quantum condition refining the usual EBK condition. This
is achieved using an old forgotten idea of Enrico Fermi.

\end{abstract}

\section{Introduction}

We address in this paper the following simple questions:

\begin{quotation}
\emph{Given a wave function }$\psi(x)$\emph{ defined on the real line, is it
possible to give an unambiguous two-dimensional pictorial representation of
that function as a surface in position-momentum phase plane? Conversely, under
which conditions can one associate to such a surface a wave function?}
\end{quotation}

We will show that there is indeed a one-to-one correspondence between wave
functions and surfaces in phase plane whose boundary is a closed curve
satisfying a certain quantum condition. Our proof is based on two results, the
first of which has almost sunk into oblivion: \textquotedblleft Fermi's
trick\textquotedblright\ \cite{Fermi} which associates to every wave function
a curve, and the existence of an exact quantization condition for such curves.
More precisely, we will see that:

\begin{itemize}
\item Every twice differentiable complex function $\psi$ satisfies a linear
(second order) differential equation%
\begin{equation}
\left[  \left(  -i\hbar\frac{d}{dx}-f(x)\right)  ^{2}+g(x)\right]  \psi=0;
\label{eq1}%
\end{equation}

\item Closed phase plane curves
\begin{equation}
p=p^{+}(x)\geq0\text{ , }p=p^{-}(x)\leq0\text{ \ }(x_{A}\leq x\leq
x_{B})\label{cur1}%
\end{equation}
are in one-to-one correspondence with functions $\psi$ provided they satisfy
an \emph{exact} quantization condition \cite{maxu1,maxu2,qidong}.
\end{itemize}

\noindent\textbf{Acknowledgements}. The present work has been supported by the
grant P20442-N13 of the Austrian Research Agency FWF.

\section{Fermi's trick}

In a largely forgotten paper \cite{Fermi} from 1930 Enrico Fermi shows that
one could associate to every quantum state $\psi$ a certain curve
$g_{\mathrm{F}}(x,p)=0$ in phase plane. He remarked that one could explicit
determine a second order differential equation satisfied by a twice
differentiable complex function. Fermi's work has recently rediscovered by
Benenti and Strini \cite{best,benenti}. The underlying idea is actually
surprisingly simple. It consists in observing that any complex twice
continuously differentiable function $\psi(x)=R(x)e^{iS(x)/\hslash}$ ($R(x)>0$
and $S(x)$ real) defined on configuration space trivially satisfies the
partial differential equation
\begin{equation}
\left(  -\hbar^{2}\frac{d^{2}}{dx^{2}}+\hbar^{2}\frac{R^{\prime\prime}}%
{R}\right)  R=0\label{trivial}%
\end{equation}
(it is assumed throughout that $R$ satisfies the concavity condition
$R^{\prime\prime}\leq0$). Performing the gauge transformation%
\[
-i\hbar\frac{d}{dx}\longrightarrow-i\hbar\frac{d}{dx}-S^{\prime}%
\]
this equation is equivalent to
\begin{equation}
\left[  \left(  -i\hbar\frac{d}{dx}-S^{\prime}\right)  ^{2}+\hbar^{2}%
\frac{R^{\prime\prime}}{R}\right]  \psi=0\label{gf1}%
\end{equation}
(the equivalence of Eqns. (\ref{trivial}) and (\ref{gf1}) can also be verified
by a direct explicit calculation). The operator
\begin{equation}
\widehat{g_{\mathrm{F}}}=\left(  -i\hbar\frac{d}{dx}-S^{\prime}\right)
^{2}+\hbar^{2}\frac{R^{\prime\prime}}{R}\label{fermop}%
\end{equation}
appearing in the left-hand side of Eqn. (\ref{gf1}) is the quantization of the
real function
\begin{equation}
g_{\mathrm{F}}(x,p)=\left(  p-S^{\prime}\right)  ^{2}+\hbar^{2}\frac
{R^{\prime\prime}}{R}\label{gf2}%
\end{equation}
and the equation $g_{\mathrm{F}}(x,p)=0$ in general determines a curve
$\mathcal{C}_{\mathrm{F}}$ in the phase plane, which Fermi ultimately
\emph{identifies} with the state $\psi$ itself. Notice that the Fermi function
only depends on the state\ $\psi$ in the sense that if the replacement of
$\psi$ with $\lambda\psi$ ($\lambda\neq0$ a complex constant) changes neither
$g_{\mathrm{F}}$ nor $\widehat{g_{\mathrm{F}}}$.

Let us illustrate Fermi's trick when $\psi$ is a Gaussian $\psi_{a,b}%
(x)=e^{-(a+ib)x^{2}/2\hbar}$. In this case we have $S(x)=-bx^{2}/2$ and
$R(x)=e^{-ax^{2}/2\hbar}$ hence%
\[
g_{\mathrm{F}}(x,p)=(p+bx)^{2}+ax^{2}-a\hbar\text{.}%
\]
The Gaussian $\psi_{a,b}$ is thus a solution of the eigenvalue problem%
\[
\frac{1}{2}\left[  \left(  -i\hbar\frac{d}{dx}+bx\right)  ^{2}+ax^{2}\right]
\psi=\frac{1}{2}a\hbar\psi.
\]
For instance, if $a=1$ and $b=0$ one recovers the fact that the standard
Gaussian $e^{-x^{2}/2\hbar}$ is the ground state of the harmonic oscillator.

\section{An exact quantization rule}

Consider the one-dimensional stationary Schr\"{o}dinger equation%
\begin{equation}
-\frac{\hbar^{2}}{2m}\frac{d^{2}}{dx^{2}}\psi(x)=[E-V(x)]\psi(x).
\label{schstat1}%
\end{equation}
We assume that the potential $V$ is piecewise continuous, and that there exist
exactly two real values $x_{A}$ and $x_{B}$ (\textquotedblleft turning
points\textquotedblright) such that%
\begin{align*}
V(x)  &  >E\text{ \ \textit{for} \ }-\infty<x<x_{A}\text{ \ or \ }%
x_{B}<x<+\infty\\
V(x)  &  =E\text{ \ \textit{for} }x=x_{A}\text{ \ or }x=x_{B}\\
V(x)  &  <E\text{ \ \textit{for} }x_{A}<x<x_{B}.
\end{align*}
It follows from elementary functional analysis (Sturm--Liouville theory) that
the equation (\ref{schstat1}) has non-zero solutions only for a set of
discrete values $E_{0}\leq E_{1}\leq\cdot\cdot\cdot$ of the energy $E$. In
\cite{maxu1,maxu2} Ma and Xu have shown that these values can be explicitly
calculated using an exact quantization rule. The argument goes as follows: let
$\chi=\psi^{\prime}/\psi$ be the logarithmic derivative of the wavefunction
$\psi$; a straightforward calculation shows that Schr\"{o}dinger's equation
(\ref{schstat1}) is equivalent to the Riccati equations%
\begin{align*}
-\chi^{\prime}(x)  &  =k(x)^{2}+\chi(x)^{2}\text{ \ \textit{for} \ }V(x)\leq
E\\
-\chi^{\prime}(x)  &  =-k(x)^{2}+\chi(x)^{2}\text{ \ \textit{for} \ }V(x)\geq
E;
\end{align*}
where $k(x)$ is given by%
\begin{equation}
k(x)=\left\{
\begin{array}
[c]{c}%
\dfrac{1}{\hbar}\sqrt{2m(E-V(x))}\text{ \ \textit{for} \ }V(x)\leq E\medskip\\
\dfrac{1}{\hbar}\sqrt{2m(V(x)-E)}\text{ \ \textit{for} \ }V(x)\geq E
\end{array}
\right.  \label{hk1}%
\end{equation}
(the function $p(x)=\hbar k(x)$ is the momentum ). Supposing that the
potential $V(x)$ is continuous at the turning points $x_{A}$ and $x_{B}$, the
exact quantum condition of Ma and Xu is then%
\begin{equation}
\int_{x_{A}}^{x_{B}}k(x)dx-\int_{x_{A}}^{x_{B}}\chi(x)[\chi^{\prime}%
(x)]^{-1}k^{\prime}(x)dx=N\pi\label{QR1}%
\end{equation}
where $N$ is the number of nodes of $\chi$ for $V(x)\leq E$ (hence $N-1$ is
the number of nodes of $\psi$ in that region). Notice that if the second
integral is neglected, the formula above reduces to the familiar approximate
Bohr--Sommerfeld or WKB prescription%
\begin{equation}
\int_{x_{A}}^{x_{B}}k(x)dx=N\pi. \label{wkb1}%
\end{equation}
We mention that Barclay \cite{barclay} had shown in some older and
unfortunately rather overlooked work that one can obtain exact quantization
rules by a clever re-summation procedure of the higher order terms in the WKB series.

A crucial fact noted by Qian and Dong \cite{qidong} (also see Serrano
\textit{et al}. \cite{serrano,serrano2}) is that condition (\ref{QR1}) can be
put for all exactly solvable systems in the simple form%
\begin{equation}
\int_{x_{A}}^{x_{B}}k(x)dx-\int_{x_{A}}^{x_{B}}k_{0}(x)dx=(N-1)\pi\label{QR2}%
\end{equation}
where the function $k_{0}$ is defined by (\ref{hk1}) with $E=E_{0}$ (the
ground energy level):
\begin{equation}
\hbar k_{0}(x)=\sqrt{2m(E_{0}-V(x))}\text{ \ \textit{for} \ }V(x)\leq E.
\label{hk0}%
\end{equation}
The exact quantization condition thus becomes%
\begin{equation}
\int_{x_{A}}^{x_{B}}\sqrt{2m[E-V(x)]}dx=\int_{x_{A}}^{x_{B}}\sqrt
{2m[E_{0}-V(x)]}dx+\frac{Nh}{2} \label{QR3}%
\end{equation}
which shows that the energy levels $E_{1},E_{2},...$ are determined by the
ground state energy level $E_{0}$. Denoting by $\mathcal{C}$ (resp.
$\mathcal{C}_{0}$) the curve
\[
\frac{1}{2m}p^{2}+V(x)=E\text{ \ (resp. }\frac{1}{2m}p^{2}+V(x)=E_{0}\text{)}%
\]
this is equivalent to
\begin{equation}
\int\nolimits_{\mathcal{C}}pdx=\int\nolimits_{\mathcal{C}_{0}}pdx+Nh.
\label{cpdx}%
\end{equation}

While a general condition for systems with $n$ degrees of freedom is still
lacking, the procedure outlined above still applies to $n$-dimensional systems
with spherical symmetry in which case $V(x)$ is replaced by the effective
potential $V_{\mathrm{eff}}(r)$ (see \cite{gu,maxu1,maxu2,qidong}). This
allows, in particular, to recover the energy levels of the hydrogen atom.

\section{Wave-functions and surfaces in phase plane}

\subsection{From $\psi$ to $\Omega$}

Let $\psi=Re^{iS/\hbar}$ ($R>0$) be a solution of the stationary
Schr\"{o}dinger equation (\ref{schstat1}); assume for instance it corresponds
to the $N$-th energy level $E_{N}$. A direct calculation shows that the
function $R$ satisfies the equation%
\begin{equation}
\frac{1}{2m}\left(  -i\hbar\frac{d}{dx}-S^{\prime}(x)\right)  ^{2}%
R(x)=[E-V(x)]R(x) \label{a}%
\end{equation}
and the corresponding Fermi function is thus%
\begin{equation}
g_{\mathrm{F}}(x,p)=(p-S^{\prime}(x))^{2}+2m(E_{N}-V(x)). \label{gf}%
\end{equation}
The area of the surface $\Omega$ bounded by the curve $\mathcal{C}%
:g_{\mathrm{F}}(x,p)=0$ is given by the formula%
\[
\operatorname{Area}(\Omega)=2\int_{x_{A}}^{x_{B}}\sqrt{2m[E_{N}-V(x)]}dx.
\]
In view of the quantization condition (\ref{QR3}) this means that we have%
\begin{equation}
\operatorname{Area}(\Omega)=Nh+2\int_{x_{A}}^{x_{B}}\sqrt{2m[E_{0}-V(x)]}dx.
\label{area1}%
\end{equation}

Let us illustrate this on the harmonic oscillator with classical Hamiltonian%
\[
H=\frac{1}{2m}(p^{2}+m^{2}\omega^{2}x^{2}).
\]
The ground energy is $E_{0}=\frac{1}{2}\hbar\omega$ and the corresponding
turning points are $\pm\sqrt{\hbar/m\omega}$ hence
\[
\operatorname{Area}(\Omega)=Nh+2\sqrt{m\hbar\omega}\int_{-\sqrt{\hbar/m\omega
}}^{\sqrt{\hbar/m\omega}}\sqrt{1-\frac{m\omega}{\hbar}x^{2}}dx;
\]
the integral is easily evaluated, and one finds that%
\begin{equation}
\operatorname{Area}(\Omega)=(N+\tfrac{1}{2})h \label{nh1}%
\end{equation}
and one thus recovers the fact that the action increases by jumps equal to $h$
starting from the initial value $\tfrac{1}{2}h$.

\subsection{From $\Omega$ to $\psi$}

We now address the converse problem. Consider a smooth closed curve
$\mathcal{C}$ (see Fig. 1) described by equations%
\begin{equation}
p=p^{+}(x)\text{ , }p=p^{-}(x)\text{\ \ \ \ \ }(x_{A}\leq x\leq x_{B});
\end{equation}
with $p^{+}(x_{A})=$ $p^{-}(x_{A}),$ $p^{+}(x_{B})=$ $p^{-}(x_{B})$ and
$p^{+}(x)\geq0$, $p^{+}(x)\leq0$. Defining functions $f(x)$ and $g(x)$
through
\begin{equation}
p^{+}(x)=f(x)+\sqrt{-g(x)}\text{ \ , \ }p^{-}(x)=f(x)-\sqrt{-g(x)} \label{pf}%
\end{equation}
the curve $\mathcal{C}$ is given by the single equation
\begin{equation}
(p-f(x))^{2}+g(x)=0 \label{f1}%
\end{equation}
and the area of the surface $\Omega$ enclosed by $\mathcal{C}$ is
\begin{equation}
\operatorname{Area}(\Omega)=2\int_{x_{A}}^{x_{B}}\sqrt{-g(x)}dx. \label{areg}%
\end{equation}

We are going to show that if $\mathcal{C}$ is quantized in the sense above, we
can associate to it a solution $\psi(x)=Re^{iS(x)/\hbar}$ of some stationary
Schr\"{o}dinger equation. Choose a number $E>0$ and define two functions
$S(x)$ and $V(x)$ by%
\begin{equation}
f(x)=S^{\prime}(x)\text{ \ , \ }g(x)=2m(V(x)-E) \label{ffi1}%
\end{equation}
that is
\begin{equation}
S(x)=\int_{x_{0}}^{x}f(x^{\prime})dx^{\prime}\text{ \ , \ }V(x)=\frac
{g(x)}{2m}+E \label{ffi2}%
\end{equation}
where $x_{0}$ is an arbitrary fixed number. Eqn. (\ref{f1}) for $\mathcal{C}$
is then
\begin{equation}
\frac{1}{2m}(p-S^{\prime}(x))^{2}+V(x)-E=0. \label{eqfermi}%
\end{equation}
Consider the differential operator
\[
\frac{1}{2m}\left(  -i\hbar\frac{d}{dx}-S^{\prime}(x)\right)  ^{2}+V(x);
\]
it has a discrete spectrum consisting of positive eigenvalues. Let $E_{0}$ be
the smallest eigenvalue; the boundary of $\Omega$ must thus satisfy the
quantization condition (\ref{QR3}), that is
\[
\int_{x_{A}}^{x_{B}}\sqrt{2m[E-V(x)]}dx=\int_{x_{A}}^{x_{B}}\sqrt
{2m[E_{0}-V(x)]}dx+\frac{Nh}{2}.
\]
This condition determines the value $E$. Let us now look for a function $R(x)$
such that
\begin{equation}
V(x)-E=\frac{\hbar^{2}}{2m}\frac{R^{\prime\prime}(x)}{R(x)} \label{r}%
\end{equation}
that is%
\begin{equation}
-\frac{\hbar^{2}}{2m}R^{\prime\prime}(x)+(V(x)-E)R(x)=0. \label{rbis}%
\end{equation}
The function $\psi(x)=Re^{iS(x)/\hbar}$ is a solution of the equation%
\[
\left[  \frac{1}{2m}\left(  -i\hbar\frac{\partial}{\partial x}-S^{\prime
}(x)\right)  ^{2}+V(x)\right]  \psi(x)=0
\]
as is shown by a direct calculation using the identity (\ref{rbis}); $\psi(x)$
is, therefore, the wave-function we are looking for. Notice that the choice of
the value $x_{0}$ in (\ref{ffi2}) is irrelevant because if we replace it with
another value $x_{0}^{\prime}\neq x_{0}$ it changes $\psi$ into $\psi^{\prime
}=e^{i\gamma/\hbar}\psi$ where $\gamma=\int_{x_{0}}^{x_{0}^{\prime}%
}f(x^{\prime})dx^{\prime}$ is a constant phase.

\section{Examples}

\subsection{Squeezed states}

We consider (unnormalized) squeezed coherent states%
\begin{equation}
\psi_{a,b}(x)=e^{-\frac{1}{2\hbar}(a+ib)x^{2}} \label{coh1}%
\end{equation}
where $a$ and $b$ are real and $a>0$. The functions $S(x)=-\frac{1}{2}bx^{2}$
and $R(x)=e^{-ax^{2}/2\hbar}$ satisfy
\begin{equation}
S^{\prime}(x)=-bx\text{ \ , \ }\frac{R^{\prime\prime}(x)}{R(x)}=-\frac
{a}{\hbar}+\frac{1}{\hbar^{2}}a^{2}x^{2}. \label{tr}%
\end{equation}
The Fermi function of $\psi_{a,b}$ is thus the quadratic form
\begin{equation}
g_{\mathrm{F}}(x,p)=(p+bx)^{2}+a^{2}x^{2}-a\hbar. \label{gf3}%
\end{equation}
Setting $z=%
\begin{pmatrix}
x\\
p
\end{pmatrix}
$ we can rewrite formula (\ref{gf3}) as
\[
g_{\mathrm{F}}(x,p)=z^{T}Mz-a\hbar
\]
where $M$ is the symmetric matrix
\begin{equation}
M=%
\begin{pmatrix}
a^{2}+b^{2} & b\\
b & 1
\end{pmatrix}
. \label{mf}%
\end{equation}
Since $\det M=a^{2}$ it follows that the surface $\Omega$ enclosed by the
ellipse $\mathcal{C}:g_{\mathrm{F}}(x,p)=0$ is $\tfrac{1}{2}h$.

A straightforward calculation shows that the matrix (\ref{mf}) factorizes as%
\begin{equation}
M=S^{T}%
\begin{pmatrix}
a & 0\\
0 & a
\end{pmatrix}
S \label{mfs}%
\end{equation}
where $S$ is the unimodular matrix
\begin{equation}
S=%
\begin{pmatrix}
a^{1/2} & 0\\
a^{-1/2}b & a^{-1/2}%
\end{pmatrix}
. \label{ess}%
\end{equation}
It turns out --and this is really a striking fact!-- that $M_{\mathrm{F}}$ is
closely related to the Wigner transform
\begin{equation}
W\psi_{a,b}(z)=\frac{1}{2\pi\hbar}\int_{-\infty}^{\infty}e^{-\frac{i}{\hbar
}py}\psi_{a,b}(x+\tfrac{1}{2}y)\psi_{a,b}^{\ast}(x-\tfrac{1}{2}y)dy
\label{oupsi}%
\end{equation}
of the state $\psi_{a,b}$. In fact (see e.g. de Gosson \cite{Birk}, Littlejohn
\cite{Littlejohn}),%
\begin{equation}
W\psi_{a,b}(z)=(\pi\hbar)^{-1/2}a^{-1/2}e^{-z^{T}Gz/\hbar} \label{goupsi}%
\end{equation}
where $G$ is the matrix%
\begin{equation}
G=S^{T}S=%
\begin{pmatrix}
a+b^{2}/a & b/a\\
b/a & 1/a
\end{pmatrix}
. \label{G}%
\end{equation}
It follows from Eqn. (\ref{mfs}) that%
\begin{equation}
W\psi_{a,b}(z)=(\pi\hbar)^{-1/2}(\det a)^{-1/2}e^{-a}\exp\left[  -\frac
{1}{\hbar}g_{\mathrm{F}}(S^{-1}D^{-1/2}Sz)\right]  \label{wgf}%
\end{equation}
with $D=%
\begin{pmatrix}
a & 0\\
0 & a
\end{pmatrix}
$.

In particular, when $n=1$ and $\psi_{a,b}(x)=e^{-x^{2}/2\hbar}$ we have
$S^{-1}D^{-1/2}S=I$ and $a=1$ and hence
\[
W\psi(z)=\left(  \pi\hbar\right)  ^{-1/4}e^{-1}e^{-\frac{1}{\hbar}z^{T}Mz}%
\]
which was already noticed by Benenti and Strini \cite{best}.

\subsection{Hermite functions}

The $N$-th Hermite function is%
\begin{equation}
h_{N}(x)=(\sqrt{\pi}2^{N}N!)^{-1/2}e^{-Q^{2}/2}H_{N}(Q)\text{ \ , \ }Q=\left(
\frac{m\omega}{\hbar}\right)  ^{1/2}x \label{herm1}%
\end{equation}
where
\begin{equation}
H_{N}(x)=(-1)^{N}e^{x^{2}}\frac{d^{N}}{dx^{N}}e^{-x^{2}} \label{herm2}%
\end{equation}
is the $N$-th Hermite polynomial; the latter satisfies the second-order
differential equation
\begin{equation}
H_{N}^{\prime\prime}(x)-2xH_{N}^{\prime}(x)+2nH_{N}(x)=0. \label{herm3}%
\end{equation}
Here, $S(x)=0$ and\ a straightforward calculation using the relation
(\ref{herm3}) yields
\begin{equation}
\frac{R^{\prime\prime}(x)}{R(x)}=\frac{m^{2}\omega^{2}}{\hbar^{2}}%
x^{2}-(2N+1)\frac{m\omega}{\hbar}; \label{herm4}%
\end{equation}
hence the Fermi function is here%
\begin{equation}
g_{\mathrm{F}}(x,p)=p^{2}+m^{2}\omega^{2}x^{2}-(2N+1)m\omega\hbar.
\label{fermiherm}%
\end{equation}
The curve $\mathcal{C}:g_{\mathrm{F}}(x,p)=0$ is again an ellipse, enclosing a
surface $\Omega$ with area%
\begin{equation}
\operatorname{Area}(\Omega)=(N+\tfrac{1}{2})h. \label{nh2}%
\end{equation}

This example is not very instructive, because the Fermi operator
$\widehat{g}_{\mathrm{F}}$ is, up to the factor $1/2m$ just $\widehat{H}%
-(N+\frac{1}{2})\hbar\omega$ where%
\[
\widehat{H}=-\frac{\hbar^{2}}{2m}\frac{d^{2}}{dx^{2}}+\frac{1}{2}m\omega
^{2}x^{2}%
\]
is the harmonic oscillator Hamiltonian, whose eigenstates are precisely the
Hermite functions (\ref{herm1}). We leave it to the reader to verify that the
same situation occurs for every real function which is an eigenstate of some
arbitrary operator
\[
\widehat{H}=-\frac{\hbar^{2}}{2m}\frac{d^{2}}{dx^{2}}+V.
\]

\end{document}